\newcommand{\diracslash}[1]{#1\llap{/\kern2pt}}

\newcommand{\be}{\begin{equation}}
\newcommand{\ee}{\end{equation}}
\newcommand{\bea}{\begin{eqnarray}}
\newcommand{\eea}{\end{eqnarray}}
\newcommand{\ba}[1]{\begin{array}{#1}}
\newcommand{\ea}{\end{array}}

\newcommand{\bt}{\begin{tabular}}
\newcommand{\et}{\end{tabular}}

\newcommand{\beas}{\begin{eqnarray*}}
\newcommand{\eeas}{\end{eqnarray*}}

\documentclass[preprint,prd,aps,floats,nofootinbib,floatfix]{revtex4}
\usepackage{graphicx}
\begin{document}

%\date{\today}
\title{Masses and Decay widths of Charmonium states 
in presence of strong magnetic fields}
%%%%%%%%%%%%%%%%%%%%%%%%%%%%%%%%%%%%%%%%%%%%%%%%%%%%%%%%%%%%%%%%%
\author{Amruta Mishra}
\email{amruta@physics.iitd.ac.in}
\affiliation{Department of Physics, Indian Institute of Technology, Delhi,
Hauz Khas, New Delhi -- 110 016, India}

\author{S.P. Misra}
\email{misrasibaprasad@gmail.com}
\affiliation{Institute of Physics, Bhubaneswar -- 751005, India} 

\begin{abstract}
The masses and decay widths of charmonium states are studied 
in the presence of strong magnetic fields. The mixing between 
the pseudoscalar and vector charmonium states at rest is observed 
to lead to appreciable negative (positive) shifts in the masses 
of the pseudoscalar (longitudinal component of the vector) 
charmonium states in vacuum/hadronic medium
in the presence of high magnetic fields. 
The pseudoscalar and vector charmonium masses in the hadronic medium,
calculated in an effective chiral model from the medium changes 
of a scalar dilaton field, have additional significant modifications 
due to the mixing effects.
The masses of the $D$ and $\bar D$ mesons in the magnetized hadronic matter
are calculated within the chiral effective model. 
The partial decay widths of the vector charmonium state 
to $D\bar D$ are computed using a field theoretical model 
for composite hadrons with quark/antiquark constituents, 
and are compared to the decay widths calculated 
using an effective hadronic Lagrangian. 
The effects of the mixing are observed to lead to significant 
contributions to the masses of the pseusoscalar and vector charmonium 
states, and an appreciable increase in the decay width 
$\psi(3770) \rightarrow D\bar D$ at large values of the magnetic fields. 
These studies of the charmonium states in strong magnetic fields
should have observable consequences on the dilepton spectra, 
as well as on the production of the open 
charm mesons and the charmonium states in ultra relativistic
heavy ion collision experiments.
\end{abstract}

\maketitle

\def\bfm#1{\mbox{\boldmath $#1$}}
\def\bfs#1{\mbox{\bf #1}}

\section{Introduction}
The study of properties of hadrons under extreme conditions,
e,g, high temperatures and/or densities is a topic of
intense research in strong interaction physics. This is due 
to its relevance in the ultra relativistic heavy ion collision 
experiments at various high energy particle accelerators,
as well as in the study of the bulk matter of
astrophysical objects, e.g., neutron stars.
In the recent past, there have been a lot of studies
on the in-medium properties of the heavy flavour mesons 
\cite{Hosaka_Prog_Part_Nucl_Phys}, 
as these can have observable consequences in high energy 
heavy ion collision experiments. 
The heavy quarkonium (charmonium and bottomonium) states
have been studied using the potential models 
\cite{eichten_1,eichten_2,Klumberg_Satz_Charmonium_prod_review,Quarkonia_QGP_Mocsy_IJMPA28_2013_review,repko}, 
the QCD sum rule approach
\cite{open_heavy_flavour_qsr,kimlee,klingl,amarvjpsi_qsr},  
the Quark Meson cooupling (QMC) model
\cite{QMC_Krein_Thomas_Tsushima_Prog_Part_Nucl_Phys_2018}, 
the coupled channel approach \cite{tolos_heavy_mesons}, 
and a chiral effective model \cite{amarvdmesonTprc,amarvepja}. 
The heavy quarkonium (charmonium and bottomonium) state 
in the presence of a gluon field has been studied in 
Refs. \cite{pes1,pes2,voloshin}. 
Assuming the heavy quark ($Q$) and antiquark ($\bar Q$) 
in the quarkonium state to be bound by color Coulomb 
potential, and the $Q\bar Q$ separation to be small compared to the
scale of the gluonic fluctuations, the mass shift in the
quarkonium state in the leading order is observed to be proportional 
to to the medium modifications of the scalar gluon condensate.  
Using the leading order formula and the linear density approximation 
for the gluon condensate
in the nuclear medium, the mass modifications of the charmonium 
states have been studied in Ref. \cite{leeko}.
The  in-medium masses of the charmonium states have been computed 
within a chiral effective model \cite{amarvdmesonTprc,amarvepja},
from the medium change of a dilaton scalar field,
which simulates the gluon condensate within the hadronic model.

The magnetic fields created in non-central 
ultra relativistic heavy ion collision experiments,
have been estimated to be huge, e.g., 
$eB \sim 2 m_\pi^2$ at Relativistic Heavy Ion Collider (RHIC) at BNL
and $eB \sim 15 m_\pi^2$ at Large Hadron Collider (LHC) at CERN 
\cite{Tuchin_Review_Adv_HEP_2013}. This
has led to a lot of work
on the study of effects of strong magnetic fields
on the properties of the hadrons.
The strong magnetic fields produced in the high energy 
heavy ion collisons rapidly drop after the collision.
This leads to induced currents, which slow down the decrease
in the magnetic field. The time evolution of the magnetic field
\cite{Tuchin_Review_Adv_HEP_2013} is still an open question,
and needs the solutions of the magnetohydrodynamic equations,
with a proper estimate of the electrical conductivity of the medium.
In the (near) central collisions, the impact parameter is small,
the magnetic field produced is weak and the created 
medium is dense. On the other hand, strong magnetic feilds 
are created in the peripheral ultra relativistic heavy 
ion collisions and the formed medium has low density. 
The effects of the magnetic fields on the heavy quarkonium
(charmonium and bottomonium) states can have observable 
consequences as these are formed in heavy ion collisions
at the early stage, when the magnetic fields can still be 
large. The open charm mesons 
\cite{Gubler_D_mag_QSR,machado_1,B_mag_QSR,dmeson_mag}
as well as the charmonium states
\cite{charmonium_mag,charmonium_mag_QSR,charmonium_mag_lee}
have been studied in the presence of magnetic fields. 
There is mixing of the pseudoscalar and vector mesons
in the presence of magnetic fields, which modifies the
properties of the charmonium states 
\cite{charmonium_mag_QSR,charmonium_mag_lee,Alford_Strickland_2013,Suzuki_Lee_2017}.
The masses of the charmonium states in the presence of magnetic fields 
have been studied in a consistent manner using QCD sum rule approach,
incorporating the mixing of the pseudoscalar and vector charmonium states 
in the hadron spectral function on the phenomenological side, as well
as, including the effects of magnetic field on the OPE
(operator product expansion) side 
\cite{charmonium_mag_QSR,charmonium_mag_lee}.
The QCD sum rule approach \cite{charmonium_mag_QSR,charmonium_mag_lee}, 
as well as, a study of the charm-anticharm bound state described
by an effective potential and solving the Schrodinger equation
in the presence of an external magnetic field
\cite{Alford_Strickland_2013} show that 
the charmonium masses have dominant contributions
from the mixing effects. A study of the mixing effects
on the formation time of the charmonia are observed
to lead to delayed (faster) formation time of the vector 
(pseudoscalar) charmonium states \cite{Suzuki_Lee_2017}.
The $J/\psi-\eta_c$ as well as $\psi'-\eta_c'$ mixings 
and the faster formation time of the pseudoscalar mesons,
might show as peaks in the dilepton spectra, 
due to `anomalous' decay modes, e.g.,
$\eta_c,\eta_c'\rightarrow l^+l^-$ 
and can act as a probe of the existence of strong 
magnetic field at the early stage \cite{Suzuki_Lee_2017}.

The masses of vector charmonium states in a (magnetized)
hadronic medium have been studied using a chiral effective model
\cite{amarvdmesonTprc,amarvepja,charmonium_mag}.
These are calculated from the medium modification of 
a scalar dilaton field which mimics the gluon condensates 
of QCD in the effective hadronic model. 
Within the chiral effective model, 
the modifications of the masses of the open charm 
($D$, $\bar D$) mesons in the (magnetized) hadronic medium
arise from their interactions with the baryons and scalar 
mesons, and have been studied in Refs. 
\cite{amarvdmesonTprc,amarvepja,dmeson_mag}.
The in-medium decay widths of the vector charmonium states 
to $D\bar D$ have been computed from the mass modifications of the
$D$, $\bar D$ and charmonium states in Refs. 
\cite{amarvdmesonTprc,amarvepja,friman,charm_decay_mag_3p0}
using a light quark pair creation model, namely the
$^3P_0$ model \cite{YOPR,YOPR_psi4040,3p0,3p0_1}, as well as using 
a field theoretical model with composite hadrons
with quark/antiquark constituents \cite{amspmwg}.
%%%%discussions added below in response to comment 3 of the Referee 
In the $^3P_0$ model, the decay of the charmonium state to $D\bar D$ 
proceeds with creation of a light quark antiquark pair in $^3P_0$ state
and the light quark (antiquark) combines with the charm antiquark 
(quark) of the decaying charmonium state
to form the $D$ and $\bar D$ mesons 
\cite{YOPR_psi4040,friman,3p0_1}. 
The light quark pair creation model \cite{YOPR_psi4040} describes 
the observed branching ratios of the decay of $\psi(4040)$ 
to $D\bar D$, $(D\bar {D^*}+\bar D D^*)$, and $D^* \bar {D^*}$,
which can not be explained using a naive quark spin counting.
The $^3P_0$ model, where the the internal structure 
of the charmonium state as well as the open charm mesons
are taken into account, is
observed to lead to strong suppression of the decay modes 
of $\psi(4040)$ to $D\bar D$ as well as 
$(D\bar {D^*}+\bar D D^*)$, as compared to the final state
$D^* \bar {D^*}$, as has been observed experimentally.
The matrix element for the decay of charmonium state to the open charm
mesons, depends on the magnitude of the momentum, $|{\bf p}|$ 
of the outgoing meson, which is given in terms of the masses 
of the decaying and the produced particles. The decay amplitude 
for $\psi(4040)$ in the $^3P_0$ model is observed to be extremely small 
for the value of $|{\bf p}|$  corresponding to the decay modes 
$D\bar D$ and $(D\bar {D^*}+\bar D D^*)$, as compared to the final state
$D^* \bar {D^*}$ \cite{YOPR_psi4040} and leads to good agreement of the 
experimentally observed branching ratios of these modes. 
The medium dependence of the decay width of the charmonium 
state to $D\bar D$ was studied using the $^3P_0$ model 
in Ref. \cite{friman} from the mass modifications of the
$D$ and $\bar D$ mesons, which in turn modify the magnitude 
of the momentum of the $D (\bar D)$ meson. The decay widths were
computed assuming harmonic oscillator wave functions
for the charmonium as well as $D$ and $\bar D$ mesons 
\cite{3p0_1}. 
For certain values of the momentum of the $D(\bar D)$ meson,
the decay widths were observed to vanish, and 
the dependence on $|{\bf p}|$ was observed to be drastically
different from the results obtained when the internal
structure of the mesons are not taken into account
\cite{friman}. 
The open flavour strong decays in vacuum have been studied extensively
in the light sector \cite{YOPR,3p0} as well as heavy flavour mesons 
\cite{YOPR_psi4040,3p0_1} accounting for the quark (antiquark) 
substructure of the hadrons, using the light quark aniquark pair 
creation in the $^3P_0$ model. The in-medium decay widths
of the charmonium states to $D\bar D$ in hadronic matter
have been studied from the mass modifications of 
the $D$ and $\bar D$ mesons as well as of the charmonium states
calculated using a chiral effective model 
\cite{amarvdmesonTprc,amarvepja,amspmwg}. 
In the present work, the effects of the magnetic field 
on the masses of the charmonium states are investigated 
accounting for the mixing of the  pseudoscalar and vector mesons, 
and, their in-medium decay widths are studied using a field theoretic
model of composite hadrons \cite{amspmwg} and compared with the
results obtained using an effective hadronic model. 
%%%%discussions added above in response to comment 3 of the Referee 

In the present work, the masses of the pseudoscalar
($\eta_c\equiv \eta_c(1S)$ and $\eta'_c\equiv \eta_c(2S)$) 
as well as the vector charmonium 
states ($J/\psi$, $\psi(2S)\equiv \psi(3686)$ 
and $\psi(1D)\equiv\psi(3770)$) in vacuum/hadronic matter
are computed accounting for 
the mixing of the pseudoscalar and the vector 
charmonium states in the presence of strong magnetic
fields. In the hadronic medium, the charmonium masses
are calculated within the chiral effective model
in the presence of a magnetic field \cite{charmonium_mag}, 
with additional contributions from the mixing effects.
The mixing is observed to lead 
to significant modifications to these charmonium masses.
The decay widths of the charmonium states to the open charm
mesons ($D\bar D$) are computed using
a field theoretical model of composite hadrons
with quark/antiquark constituents \cite{amspmwg}.
These decay widths are compared with the results 
obtained from an effective hadronic model.

The outline of the paper is as follows. In section II,
we briefly describe the chiral effective model used to
compute the masses of the (pseudoscalar and vector) charmonium 
states as well as the open charm meson masses 
in the magnetized matter. In the presence of a magnetic field, 
the mixings between the pseudoscalar and the vector mesons
are taken into account through an effective hadronic interaction.
In section III, we describe the decay of the charmonium states
to $D\bar D$ using a model for composite hadrons
as well as an effective hadronic model. In subsection A,
we describe the field theoretical model with 
composite hadrons with quark/antiquark constituents. This model
is used in the present work to compute the partial decay widths 
of the charmonium states to $D\bar D$ in 
vacuum/hadronic matter in the presence of strong magnetic fields.
The charmonium decay widths to $D\bar D$ as calculated
using the field theoretic model for composite hadrons
are compared to the results obtained from an effective 
hadronic Lagrangian as described in subsection B.
In section IV, we discuss the results obtained in the present
investigation of the charmonium masses as well as charmonium decay widths,
accounting for the mixing of the pseudoscalar and vector
charmonium states in the presence of strong magnetic fields.
In section V, we summarize the findings of the present study.

\section{Charmonium masses in strong magnetic fields}

In this section, we investigate the mass modifications 
of the vector and pseudoscalar charmonium masses 
in the presence of strong magnetic fields.
The masses of these heavy charm quark-antiquark bound states
are calculated in the magnetized nuclear matter 
from the in-medium gluon condensates, which
are simulated by a scalar dilaton field within
a chiral effective model. The mixing of the pseudoscalar
and vector charmonium states in the presence
of an external magnetic field is studied using 
an effective Lagrangian interaction term 
\cite{charmonium_mag_lee} and is observed to be the
dominant contribution to the mass shifts for these
charmonium states.

The Lagrangian density of the chiral effective model, in the presence
of a magnetic field, is given as \cite{paper3}
\bea
{\cal L} = {\cal L}_{kin} + \sum_ W {\cal L}_{BW}
          +  {\cal L}_{vec} + {\cal L}_0 
+ {\cal L}_{scalebreak}+ {\cal L}_{SB}+{\cal L}_{mag}^{B\gamma},
\label{genlag} \eea
where, $ {\cal L}_{kin} $ corresponds to the kinetic energy terms
of the baryons and the mesons,
${\cal L}_{BW}$ contains the interactions of the baryons 
with the meson, $W$ (scalar, pseudoscalar, vector, axialvector meson),
$ {\cal L}_{vec} $ describes the dynamical mass
generation of the vector mesons via couplings to the scalar fields
and contains additionally quartic self-interactions of the vector
fields, ${\cal L}_0 $ contains the meson-meson interaction terms,
${\cal L}_{scalebreak}$ is a scale invariance breaking logarithmic
potential given in terms of a scalar dilaton field, $\chi$ and 
$ {\cal L}_{SB} $ describes the explicit chiral symmetry
breaking. The term ${\cal L}_{mag}^{B\gamma}$, describes the interacion
of the baryons with the electromagnetic field, which includes
a tensorial interaction 
$\sim \bar {\psi_i} \sigma^{\mu \nu} F_{\mu \nu} \psi_i$,
whose coefficients account for the anomalous magnetic moments
of the baryons \cite{dmeson_mag}.

The trace of the energy momentum tensor of QCD is equated 
to that of the chiral effective model to obtain a relation 
between the scalar gluon condensate 
$\left\langle  \frac{\alpha_{s}}{\pi} G_{\mu\nu}^{a} G^{ \mu\nu a}
\right \rangle $
and the dilaton field $\chi$ of the
the scale breaking term ${\cal L}_{scalebreak}$.
The in-medium masses of the charmonium states are hence
obtained from the medium changes of the dilaton field.
The dilaton field, $\chi$ is solved along with the scalar
fields  (non-strange scalar-isoscalar, field $\sigma$, 
non-strange scalar isovector, $\delta$, the strange field
$\zeta$) in the magnetized hadronic matter,
from the coupled equations of motion of these fields.
The values of the scalar fields 
are used to obtain the in-medium masses of the $D$ and $\bar D$
mesons as well as the charmonium states
\cite{dmeson_mag,charmonium_mag,charm_decay_mag_3p0}.

The mass shifts of the (pseudoscalar and vector) charmonium 
states, computed from the medium change of the dilaton field calculated 
within the chiral effective model are given as
\cite{amarvdmesonTprc,amarvepja}
\begin{equation}
\Delta m_{P,V}= \frac{4}{81} (1 - d) \int d |{\bf k}|^{2} 
\langle \vert \frac{\partial \psi (\bf k)}{\partial {\bf k}} 
\vert^{2} \rangle
\frac{|{\bf k}|}{|{\bf k}|^{2} / m_{c} + \epsilon} 
 \left( \chi^{4} - {\chi_0}^{4}\right), 
\label{masspsi}
\end{equation}
where 
\begin{equation}
\langle \vert \frac{\partial \psi (\bf k)}{\partial {\bf k}} 
\vert^{2} \rangle
=\frac {1}{4\pi}\int 
\vert \frac{\partial \psi (\bf k)}{\partial {\bf k}} \vert^{2}
d\Omega.
\end{equation}
In equation (\ref{masspsi}), the value of $d$, introduced 
in the logarithmic scale breaking term ${\cal L}_{scalebreak}$, 
is equal to $(2N_f)/(11 N_c)$
for the QCD $\beta$ function calculated at the
one loop level (with $N_c$ and $N_f$
as the number of colors and flavors), but taken as a 
parameter in the effective chiral model.
The wave functions of the charmonium states,
$\psi(\bf k)$ are assumed to be harmonic oscillator
wave functions \cite{leeko,amarvdmesonTprc,amarvepja}, 
$m_c$ is the mass of charm quark,
$m_{P,V}^{vac}$ is the vacuum mass of the pseudoscalar 
(vector) charmonium state, and $\epsilon=2m_c-m_{P,V}^{vac}$
is the binding energy of the charm-anticharm bound state.

The mixings of the pseudoscalar ($P\equiv \eta_c(1S),\, \eta_c(2S)$) 
and vector ($V\equiv J/\psi, \psi(2S), \psi (1D)$) charmonium states
are taken into account through the interaction 
\cite{charmonium_mag_lee}
\begin{equation}
{\cal L}_{PV\gamma}=\frac{g_{PV}}{m_{av}} e {\tilde F}_{\mu \nu}
(\partial ^\mu P) V^\nu,
\label{PVgamma}
\end{equation}
where $m_{av}=(m_V+m_P)/2$, $m_P$ and $m_V$ are the masses 
for the pseudoscalar and vector charmonium states,
${\tilde F}_{\mu \nu}$ is the dual electromagnetic field.
In the hadronic medium, the charmonium masses are calculated 
from the medium modification of the dilaton field, 
using equation (\ref{masspsi}) within the chiral effective model. 
In equation (\ref{PVgamma}), the coupling parameter $g_{PV}$
is fitted from the observed value of the radiative decay width, 
$\Gamma(V\rightarrow P +\gamma)$ given as
\begin{equation}
\Gamma (V\rightarrow P \gamma)
=\frac{e^2}{12}\frac{g_{PV}^2 {p_{cm}}^3}{\pi m_{av}^2},
\label{decay_VP}
\end{equation}
where, $p_{cm}=(m_V^2-m_P^2)/(2m_V)$
is the magnitude of the center of mass
momentum in the final state. 
The masses of the pseudoscalar and the longitudinal component
of the vector mesons including the mixing effects 
are given by
\begin{equation}
m^{(PV)}_{P,V^{||}}=\frac{1}{2} \Bigg ( M_+^2 
+\frac{c_{PV}^2}{m_{av}^2} \mp 
\sqrt {M_-^4+\frac{2c_{PV}^2 M_+^2}{m_{av}^2} 
+\frac{c_{PV}^4}{m_{av}^4}} \Bigg),
\label{mpv_long}
\end{equation}
where $M_+^2=m_P^2+m_V^2$, $M_-^2=m_V^2-m_P^2$ and 
$c_{PV}= g_{PV} eB$.
The effective Lagrangian term given by equation 
(\ref{PVgamma}) has been observed to lead to the
mass modifications of the longitudinal $J/\psi$ and
$\eta_c$ due to the presence of the magnetic field,
which agree extermely well with  
a study of these charmonium states using a  
QCD sum rule approach incorporating the mixing effects
\cite{charmonium_mag_QSR,charmonium_mag_lee}.

The decay widths of the vector charmonium states 
to $D\bar D$ in the presence of magnetic fields
are calculated from the mass modifications of these 
charmonium states and the open charm mesons
in the magnetized nuclear matter.
The $D$ and $\bar D$ meson masses are calculated 
within the chiral effective model from their interactions
with the nucleons and scalar mesons in the magnetized nuclear
matter. The lowest Landau level contributions are retained
for the charged $D$ and $\bar D$ mesons in the presence
of the external magnetic field.
The charmonium decay widths are calculated using a field theoretical
model of composite hadrons with quark/antiquark constituents
as well as an effective hadronic Lagrangian,
as described in the following section.

\section{Decay widths of Charmonium states to $D\bar D$} 

%%%%%%%%%%%%added below 
The in-medium  decay widths of the vector charmonium states
to $D\bar D$ in the presence of strong magnetic fields 
are computed using  field theoretic model of composite
hadrons with quark/antiquark constituents, and compared to
the results using an effective hadronic Lagrangian,
when the internal structure of these mesons are ignored. 
The models  used for the calculation of the charmonium
decay widths are described in the following.
%%%%%%%%%%%%added above 

\subsection{Model of composite hadrons with quark/antiquark constituents:}
\label{dwFT_comp_had}

We investigate the charmonium decay widths to the open charm
($D$ and $\bar D$) mesons in nuclear matter
in the presence of strong magnetic fields 
using a field theoretical model with composite hadrons
\cite{spm781,spm782,spmdiffscat}.
The model describes the hadrons comprising of 
quark and antiquark constituents. 
The constituent quark field operators of the hadron in motion
are constructed from the constituent quark field operators of
the hadron at rest, by a Lorentz boosting.
Similar to the MIT bag model \cite{MIT_bag}, 
where the quarks (antiquarks) occupy
specific energy levels inside the hadron, it is assumed 
in the present model for the composite hadrons that 
the quark/antiquark constituents carry fractions
of the mass (energy) of the hadron at rest (in motion) 
\cite{spm781,spm782}.

With explicit constructions of the charmonium 
state and the open charm mesons, the decay width is
calculated using the light quark antiquark pair creation
term of the free Dirac Hamiltonian 
for constituent quark field \cite{amspmwg}.
The relevant part of the quark pair creation term is through the
 $d\bar d (u\bar u$) creation for decay 
of the charmonium state, $\Psi$, to the final state 
$D^+D^-$($D^0 {\bar {D^0}}$). 
For $\Psi \rightarrow D({\bf p}) + {\bar D} ({\bf p'})$,
this pair creation term is given as 
\begin{equation}
{\cal H}_{q^\dagger\tilde q}({\bf x},t=0)
=Q_{q}^{(p)}({\bf x})^\dagger (-i 
%{\bf {\alpha\cdot \bigtriangledown}} 
\mbox{\boldmath $\alpha$}\cdot
\mbox{\boldmath $\bigtriangledown$} 
+\beta M_q)
%{\alp}
%\newcommand{\alp}{ \mbox{\boldmath $\alpha$}  }
{\tilde Q}_q^{(p')}({\bf x}) 
\label{hint}
\end{equation}
where, $M_q$ is the constituent mass of the light quark, 
$q=(u,d)$. The subscript $q$ of the field operators 
in equation (\ref{hint})  
refers to the fact that the light antiquark, $\bar q$ and 
light quark, $q$ are the constituents 
of the $D$ and $\bar D$ mesons with momenta ${\bf p}$ and
${\bf p'}$ respectively in the final state of the
decay of the charmonium state, $\Psi$.

The charmonium state, $\Psi$ with spin projection m, at rest is written as
\begin{equation}
|\Psi_m({\bf 0})\rangle = {\int {d {\bf k} {c_r} ^i ({\bf k})^\dagger
u_r a_m(\Psi,{\bf k})\tilde {c_s}^i (-{\bf k})v_s|vac\rangle}},
\label{Psi}
\end{equation}
where, $i$ is the color index of the charm quark/antiquark operators,
$u_r$ and $v_s$ are the two component spinors for the quark and
antiquark. The expressions for $a_m (\Psi, {\bf k})$ are 
given in terms of the wave functions (assumed to be harmonic 
oscillator type) for the charmonium states
\cite{spmddbar80,amspmwg}.

The $D(D^+,D^0)$ and $\bar D(D^-,\bar {D^0})$ states, 
with finite momenta are contructed in terms of the
constituent quark field operators, obtained 
from the quark field operators of these mesons 
at rest through a Lorentz boosting \cite{spmdiffscat}. 
These states, assuming harmonic oscillaotor wave functions, 
are explicitly given as
\begin{eqnarray}
&&|D ({\bf p})\rangle  = 
\frac{1}{\sqrt{6}}
\Big (\frac {R_D^2}{\pi} \Big)^{3/4}
\int d{\bf k} 
\exp\Big(-\frac {R_D^2 {\bf k}^2}{2}\Big)
{c_r}^{i}({\bf k}+\lambda_2 {\bf p})
^\dagger u_r^\dagger 
\tilde {q_s}^{i} 
(-{\bf k} +\lambda_1 {\bf p})v_s
d\bfs k,
\label{d}
\\
%\end{eqnarray}
%\begin{eqnarray}
&&|{\bar D} ({\bf p}')\rangle 
= \frac{1}{\sqrt{6}}
\Big (\frac {R_D^2}{\pi} \Big)^{3/4}
\int d{\bf k} 
\exp\Big(-\frac {{R_D}^2 {\bf k}^2}{2}\Big)
{q_r}^{i}({\bf k}+\lambda_1 {\bf p}')
^\dagger u_r^\dagger 
\tilde {c_s}^{i} 
(-{\bf k} +\lambda_2 {\bf p}') v_s
d\bfs k,
\label{dbar}
\end{eqnarray}
where, $q=(d,u)$ for $(D^+,D^-)$ and  $(D^0,\bar {D^0})$
respectively. 
%The harmonic oscillator strength  
%parameter for $D(\bar D)$ meson is taken
%as $R_D$=(310 MeV)$^{-1}$ so as to give the experimental
%values of the vacuum decay widths of 
%$\psi (3770)\rightarrow D\bar D$,
%and, $\psi(4040)$ to $D\bar D$, $D\bar D^*$, $D^* \bar D$ and 
%$D^*\bar D^*$ \cite{leeko}. 

In equations (\ref{d}) amd (\ref{dbar}),
$\lambda_1$ and $\lambda_2$ are the 
fractions of the mass (energy) of the $D(\bar D)$ meson
at rest (in motion), carried by the constituent 
light (d,u) antiquark (quark)
and the constituent heavy charm quark (antiquark), 
with $\lambda_1 +\lambda_2=1$.
The values of $\lambda_1$ and $\lambda_2$ are calculated 
by assuming the binding energy of the hadron 
as shared by the quark (antquark) to be inversely 
proportional to the quark (antiquark) mass
\cite{spm782}. The energies of 
the light antiquark (quark) and heavy charm quark (antiquark), 
$\omega_i=\lambda_i m_D (i=1,2)$, are assumed to be 
\cite{amspmwg,spm782}
\begin{eqnarray}
\omega_1=M_q+ \frac{\mu}{M_q}\times BE,\;\; 
\omega_2=M_c+ \frac{\mu}{M_c}\times BE,
\end{eqnarray}
where $BE=(m_D-M_c-M_q)$ is the binding energy of $D (\bar D)$
meson, with $M_c$ and $M_q$ as the masses of the
constitutent charm and light quark (antiquark), and,
$\mu$ is the reduced mass of the heavy-light quark-antiquark 
system (the $D (\bar D)$ meson), defined by $1/\mu=1/M_q+1/M_c$.
The reason for making this assumption comes from the example of
hydrogen atom, which is the bound state of the proton and the electron.
As the mass of proton is much larger as compared to the mass of the 
electron, the binding energy contribution from the electron is
$\frac{\mu}{m_e}\times BE \simeq BE$ of hydrogen atom, and the
contribution from the proton is $\frac{\mu}{m_p}\times BE$, 
which is negligible as compared to the total 
binding energy of hydrogen atom, since $m_p >> m_e$. 
With this assumption, the binding energies of the heavy-light mesons, 
e.g., $D(\bar D)$ mesons \cite{amspmwg}
and, $B(\bar B)$ mesons \cite{amspm_upsilon}, 
mostly arise from the contribution from 
the light quark (antiquark).

To compute the decay width of the charmonium state, $\Psi$
to $D\bar D$, we evaluate the matrix element of 
the light quark-antiquark pair creation part of the Hamiltonian,
between the initial charmonium state and the final state for the reaction
$\Psi \rightarrow D ({\bf p})+{\bar D}({\bf p}')$
as given by
\begin{eqnarray}
\langle D ({\bf p}) | \langle {\bar D} ({\bf p}')|
{\int {{\cal H}_{d^\dagger\tilde d}({\bf x},t=0)d{\bf x}}}
|{\Psi }_m (\vec 0) \rangle 
= \delta({\bf p}+{\bf p}')A^\Psi (|{\bf p}|)p_m,
\label{tfi}
\end{eqnarray}
where,
\begin{eqnarray}
A^{\Psi}(|{\bf p}|) = 6c_\Psi\exp[(a_\Psi {b_\Psi}^2
-R_D^2\lambda_2^2){\bf p}^2]
\cdot\Big(\frac{\pi}{a_\Psi}\Big)^{{3}/{2}}
\Big[F_0^\Psi+F_1^\Psi\frac{3}{2a_\Psi}
+F_2^\Psi\frac{15}{4a_\Psi^2}\Big].
\label{ap}
\end{eqnarray}
In the above, the parameters $a_\Psi$, $b_\Psi$ and $c_\Psi$  
are given in terms of $R_D$ and $R_\Psi$, which are
the strengths of the harmonic oscillator wave functions
for the $D(\bar D)$ and the charmonium states,
and $F_i^\Psi (i=0,1,2)$ are polynomials in $|{\bf p}|$, 
the magnitude of the momentum of the outgoing $D(\bar D)$ meson
\cite{amspmwg}.
\noindent 
With $\langle f | S |i\rangle =\delta_4 (P_f-P_i) M_{fi}$,
we have
\begin{equation}
M_{fi}=2\pi (-i A^ \Psi (|{\bf p}|)p_m.
\end{equation}
The expression of the decay width of the 
charmonium state, $\Psi$ to $D\bar D$,
as calculated in the present model for composite hadrons,
without accounting for the mixing effects,
is given by
\begin{eqnarray}
&&\Gamma(\Psi\rightarrow D({\bf p}) {\bar D} (-{\bf p}))
 \nonumber \\
&=&\gamma_\Psi^2 \frac{1}{2\pi} 
\int \delta(m_{\Psi}-p^0_{D}-p^0_{\bar D})
|M_{fi}|^2_{\rm {av}}
\cdot 4\pi |{\bfs p}_{D}|^2 d|{\bfs p}_{D}| 
\nonumber\\
&=& \gamma_\Psi^2\frac{8\pi^2}{3}|{\bf p}|^3
\frac {p^0_{D}(|{\bf p}|) p^0_{\bar D}(|{\bf p}|)}{m_{\Psi}}
A^{\Psi}(|{\bf p}|)^2
\label{gammapsiddbar}
\end{eqnarray}
In the above, $p^0_{D ({\bar D})}(|{\bf p}|)
=\big(m_{D ({\bar D})}^2+|{\bf p}|^2\big)^{{1}/{2}}$, 
and the expression for 
$A^{\Psi}(|{\bf p}|)$ is given by equation (\ref{ap}).
The parameter, $\gamma_\Psi$, in the expression for 
the charmonium decay width,
is a measure of the coupling strength
for the creation of the light quark antiquark pair,
to produce the $D\bar D$ final state. 
This parameter is adjusted to reproduce the
vacuum decay widths of $\psi(3770)$ to $D^+D^-$ and
$D^0 \bar {D^0}$ \cite{amspmwg}. 
The decay width of the charmonium state is observed to have
the dependence on the magnitude of the 3-momentum
of the produced $D(\bar D)$ meson, $|{\bf p}|$, 
as a polynomial part multiplied by an exponential 
term.  
%%%%%%%%%%modifying here%%%%%%%%%%%%%%%%%%%%%%%%%
The medium modification of the charmonium decay width
is studied due to the mass modifications of the
charmonium state, the $D$ and $\bar D$ mesons through 
$|{\bf p}|$, which is given as, 
\begin{equation}
|{\bf p}|=\Big (\frac{{m_\psi}^2}{4}-\frac {{m_D}^2+{m_{\bar D}}^2}{2}
+\frac {({m_D}^2-{m_{\bar D}}^2)^2}{4 {m_\Psi}^2}\Big)^{1/2}.
\label{pd}
\end{equation}
The expression for the charmonium decay width given by
equation (\ref{gammapsiddbar}) is for the case when the
mixing of the pseudoscalar and the longitudinal component
of the vector mesons is not taken into account,
and in equation (\ref{pd}), the masses of the charmonium and
open charm mesons are the effective masses in the
hadronic matter in the presence of a magnetic field.

When we include the mixing effect, the expression
for the decay width is given as
\begin{eqnarray}
\Gamma^{PV}(\Psi &\rightarrow & D({\bf p}) {\bar D} (-{\bf p}))
=\gamma_\Psi^2\frac{8\pi^2}{3}
\Bigg [ 
\Bigg(\frac{2}{3} |{\bf p}|^3 
\frac {p^0_D (|{\bf p}|) p^0_{\bar D}(|{\bf p}|)}{m_\Psi}
A^{\Psi}(|{\bf p}|)^2 \Bigg)
\nonumber \\
&+&\Bigg(\frac{1}{3} |{\bf p}|^3 
\frac {p^0_D(|{\bf p}|) p^0_{\bar D}(|{\bf p}|)}{m_{\Psi}^{PV}}
A^{\Psi}(|{\bf p}|)^2 \Bigg) \Big({|{\bf p}|\rightarrow |{\bf p|}
(m_\Psi = m_\Psi^{PV})}\Big)
\Bigg]. 
\label{gammapsiddbar_mix}
\end{eqnarray}
In the above, the first term corresponds to the transverse
polarizations for the charmonium state, $\Psi$, whose masses 
remain unaffected by the mixing of the 
pseudoscalar and vector charmonium
states. The second term in (\ref{gammapsiddbar_mix})
corresponds to the longitudinal component of the
charmonium state whose mass is modified due to mixing
with the pseudoscalar meson in the presence of the 
magnetic field, as given by equation (\ref{mpv_long}).

%%%%%added the subsection below
\subsection{Effective hadronic model}
\label{dwFT_had}

To calculate the decay width of the vector charmonium state 
to $D\bar D$, we use the effective interaction Lagrangian 
\begin{equation}
{\cal L}_{had}= ig_{\Psi D \bar D} \Psi^\mu
((\partial_\mu \bar D )D - \bar D  (\partial_\mu D)),
\label{eff_had}
\end{equation}
where, $D=(D^0,D^+)^T$ and $\bar D=(\bar {D^0}, D^-)$.
The above Lagrangian is motivated by the hidden gauge approach 
\cite{HGA}. It might be noted here that another approach 
of introducing the massive vector mesons 
as gauge bosons using the minimal substitution 
leads to an additional term which is a contact interaction
$\sim \Psi^\mu \Psi_\mu \bar D D$
\cite{leeko}. These methods are shown to be consistent 
when both the vector and axial vector mesons are included
\cite{Meissner_Phys_Rep}.
Using the interaction Lagrangian as given by
equation (\ref{eff_had}), the decay width of the vector
charmonium state at rest decaying to 
$D\bar D (D^+D^-,D^0 {\bar {D^0}})$ 
is obtained as
\begin{equation}
\Gamma_{\rm {had}} (\Psi \rightarrow D\bar D) 
= \frac {g_{\Psi D\bar D}^2}{6 \pi m_\Psi^2} |{\bf p}|^3,
\label{gammapsippddbar_had}
\end{equation} 
where $|{\bf p}|$ is the magnitude of the momentum 
of the outgoing $D(\bar D)$ meson,
given by equation (\ref{pd}). The coupling constant
$g_{\Psi D\bar D}$ for $\Psi\rightarrow D\bar D$
is fitted from the measured decay width in vacuum. 
When the masses of the $D$ and $\bar D$ mesons are
taken to be the same, the formula for the decay width 
reduces to the expression \cite{friman}
\begin{equation}
\Big (\Gamma_{\rm {had}} (\Psi \rightarrow D\bar D)\Big)_{\rm {approx}}
= \frac {g_{\Psi D\bar D}^2}{48 \pi m_\Psi^2}
(m_\Psi^2-4 m_D^2)^{3/2},
\label{gammapsippddbar_had_friman}
\end{equation} 
for each of the channels $\psi\rightarrow D^+D^-$ and
$\Psi\rightarrow D^0 {\bar {D^0}}$,
giving the expression for the total decay width
to be twice of the expression given by equation 
(\ref{gammapsippddbar_had_friman}) \cite{friman}.

In the presence of the magnetic field, there is mixing of the
pseudoscalar and vector mesons leading to the  masses of the
longitudinal component of the vector charmonium state, 
$\Psi$ to be modified. This leads to the
decay width of $\Psi\rightarrow D\bar D$
to be modified to

\begin{equation}
\Gamma_{had}^{PV}(\Psi \rightarrow D({\bf p}) {\bar D} (-{\bf p}))
= \frac {g_{\Psi D\bar D}^2}{6 \pi m_\Psi^2}
 \Bigg[
\frac{2}{3} \Big (|{\bf p}| (m_\Psi)\Big )^3 +
\frac{1}{3} \Big (|{\bf p}| (m_\Psi \rightarrow m_\Psi ^{PV})\Big)^3
\Bigg],
\label{gammapsiddbar_mix_had}
\end{equation}
where the first term corresponds to the contributions from the
transverse components and the ssecond term from the longitudinal
component of the charmonium state $\Psi$.

%%%%%added the subsection above

\section{Results and Discussions}

In the present work, the masses and the partial decay widths
of the charmonium states to $D\bar D$ are investigated 
in the presence of strong magnetic fields. 
These are studied taking the effects of the mixing of the pseudoscalar
and vector mesons into consideration, 
in the presence of magnetic fields. 
The created magnetic fields in the peripheral ultra-relavtistic 
heavy ion collsion experiments, e.g., at RHIC and LHC, 
are huge, and the matter resulting from the high energy collision
is of (extremely) low density. In the present work,
the effects of magnetic field on the charmonium states 
are studied at zero density as well as at the nuclear matter
saturation density in the magnetized (asymmetric) nuclear matter. 
In the nuclear medium, the masses of the charmonium states 
and the open charm mesons in the presence of a magnetic field
are calculated using a chiral effective model. 
The in-medium charmonium masses are studied in the 
magnetized nuclear matter as arising from the medium
changes in a scalar dilaton field, $\chi$,
which mimics the gluon condensates of QCD in the effective hadronic 
model \cite{charmonium_mag}.
The masses of the open charm mesons in the magnetized 
nuclear matter are modified due to their
interactions with the nucleons as well as
the scalar mesons, $\sigma$, $\zeta$,
and $\delta$ \cite{dmeson_mag}.
These scalar fields and the dilaton field $\chi$ 
are solved from their coupled equations of motion.
The charmonium and open charm meson
masses are obtained from the changes in the scalar fields 
in the isospin asymmetric nuclear matter
(with isospin asymmetry parameter, $\eta$ defined as
$\eta=(\rho_n-\rho_p)/(2\rho_B)$),
in the presence of magnetic fields using the chiral
effective model \cite{charm_decay_mag_3p0}.
The effects of the mixing of the pseudoscalar and 
longitudinal vector mesons are taken into account,
which are observed to lead to significant contributions
to the charmonium masses. These, as we shall see later, 
have appreciable
effects on the partial decay width of the $\psi(3770)$ 
to $D\bar D$ at high magnetic fields.

The mass modifications of the pseudoscalar 
($\eta_c \equiv \eta_c(1S) $ and $\eta'_c \equiv \eta_c (2S))$ 
and the vector ($J/\psi$, $\psi(2S)\equiv \psi(3686)$, 
$\psi(1D)\equiv \psi(3770)$) charmonium states
are investigated in the presence of magnetic fields.
These masses are studied accounting for the
mixing of the pseudoscalar and vector mesons 
in the presence of magnetic fields,
described by the effective interaction Lagrangian term 
given by equation (\ref{PVgamma}).
For the charmonium states at rest, the mixing effect 
leads to a drop (increase) in the mass of
the pseudoscalar meson (the longitudinal component
of the vector charmonium state) as given by
equation (\ref{mpv_long}), and these are
observed to be the dominant modifications to 
the charmonium masses. 

\begin{figure}
\hskip -0.4in
\vskip -0.4in
\includegraphics[width=18cm,height=18cm]{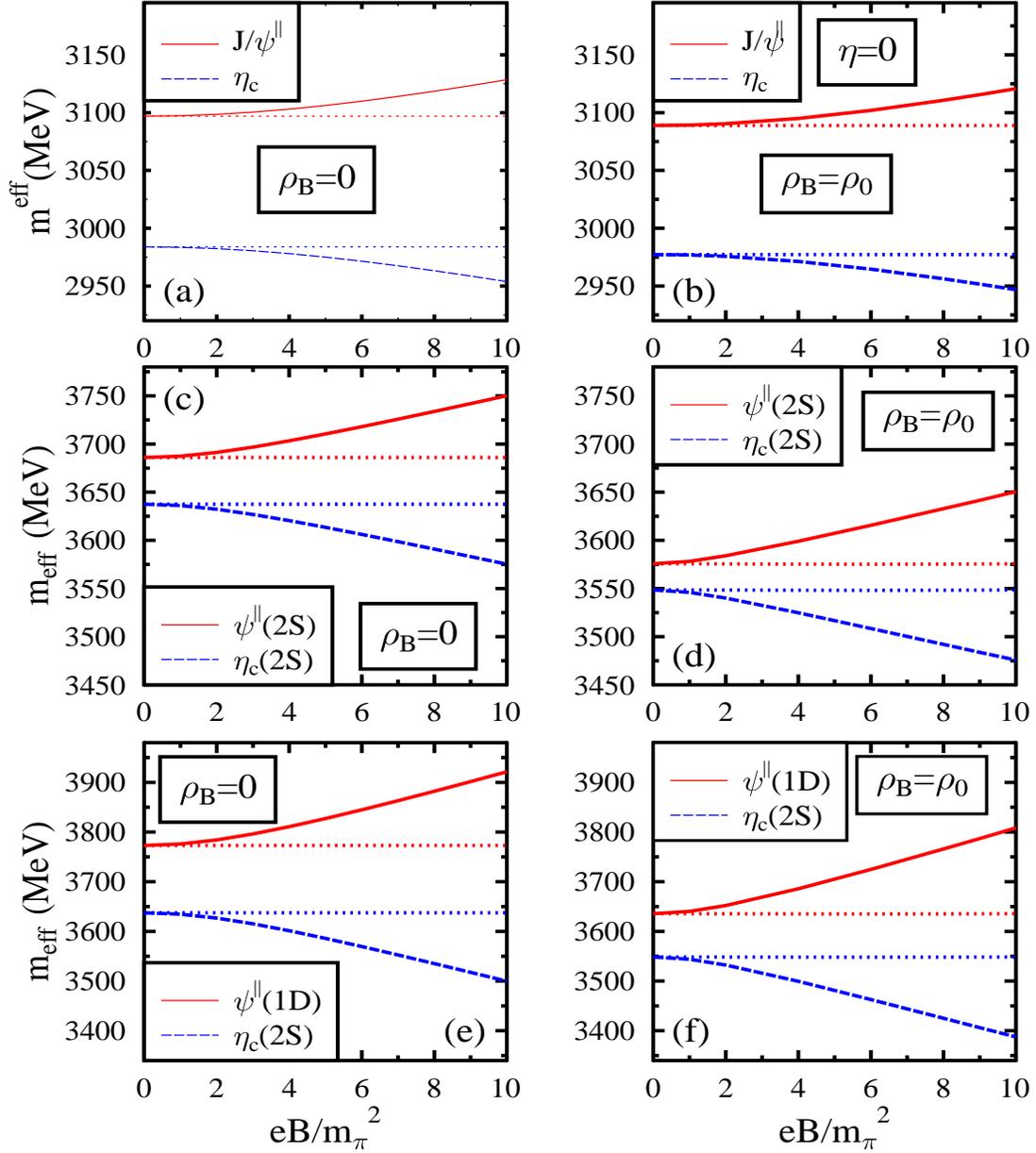}
\vskip -0.4in
\caption{(Color online)
The masses of the pseudoscalar mesons ($\eta_c \equiv \eta_c (1S)$ 
and $\eta_c' \equiv \eta_c (2S)$)
and the longitudinal components of the vector charmonium states 
($J/\psi,\psi(2S)\equiv \psi (3686), \psi(1D)\equiv  \psi(3770)$) are 
plotted as functions of $eB/{m_\pi^2}$ for $\rho_B=0$
as well as at $\rho_B=\rho_0$ in symmetric nuclear matter
($\eta$=0). The effects of the mixing between the pseudoscalar 
and the vector charmonium states ($J/\psi$ with $\eta_c$, $\psi(2S)$ 
with $\eta_c(2S)$ and $\psi(1D)$ with $\eta_c(2S)$) 
on the charmonium masses are shown and compared to the case
of not including the mixing effects (shown as dotted lines). 
}
\label{mpsi_spinmix_eta0_f}
\end{figure}

\begin{figure}
\hskip -0.4in
\vskip -0.4in
\includegraphics[width=18cm,height=18cm]{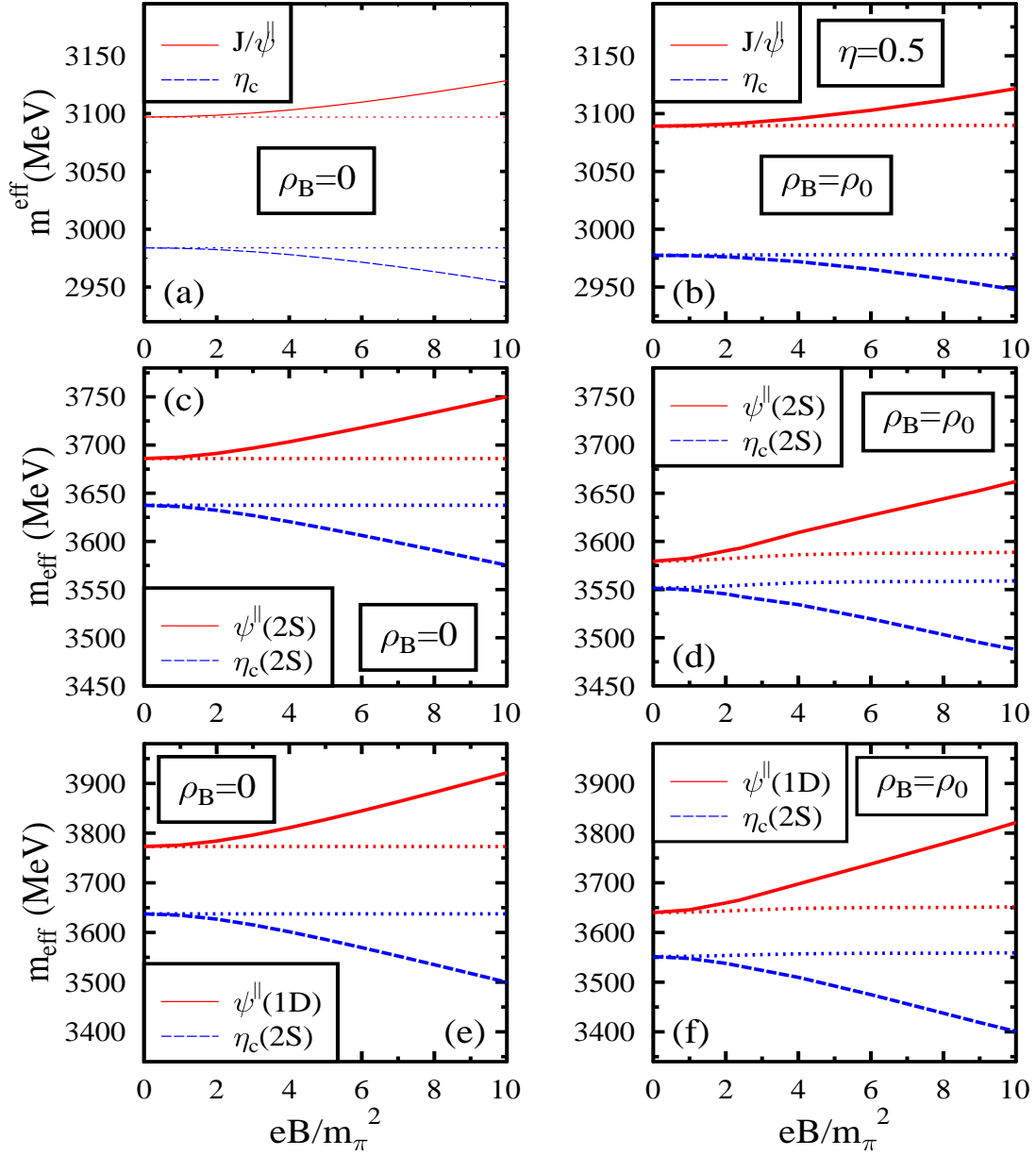}
\vskip -0.4in
\caption{(Color online)
Same as fig. \ref{mpsi_spinmix_eta0_f}, for asymmetric nuclear matter
with $\eta$=0.5.
}
\label{mpsi_spinmix_eta5_f}
\end{figure}

\begin{figure}
\hskip -0.4in
\vskip -0.4in
\includegraphics[width=18cm,height=18cm]{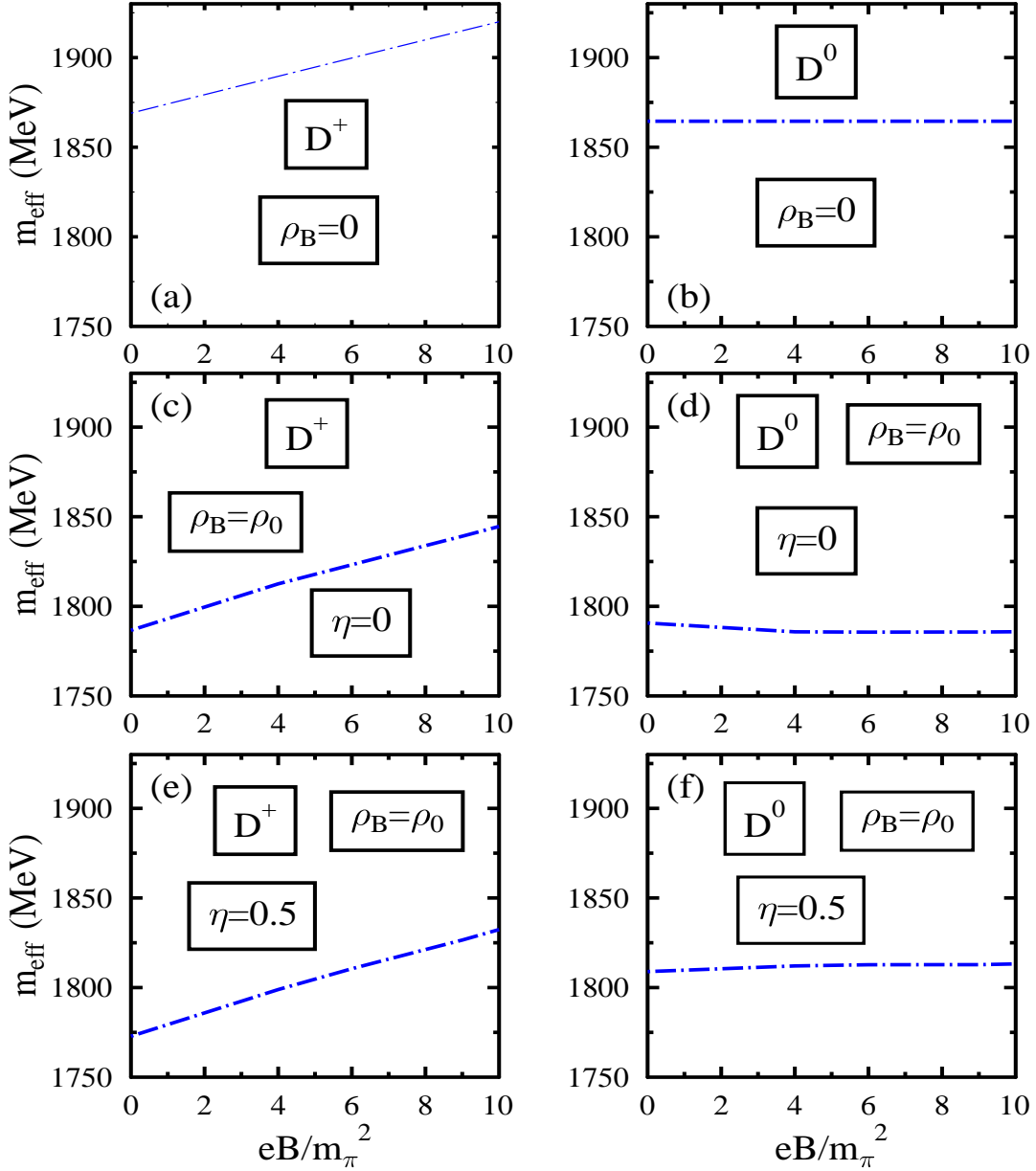}
\vskip -0.4in
\caption{(Color online)
The masses of $D$ mesons ($D^+$ and $D^0$) are plotted as functions 
of $eB/{m_\pi^2}$. These masses are shown for the case of $\rho_B=0$, 
as well as for $\rho_B=\rho_0$ in symmetric ($\eta$=0) and asymmetric 
(with $\eta$=0.5) nuclear matter.
}
\label{dmeson_mag_rh0_f}
\end{figure}

\begin{figure}
\hskip -0.4in
\vskip -0.4in
\includegraphics[width=18cm,height=18cm]{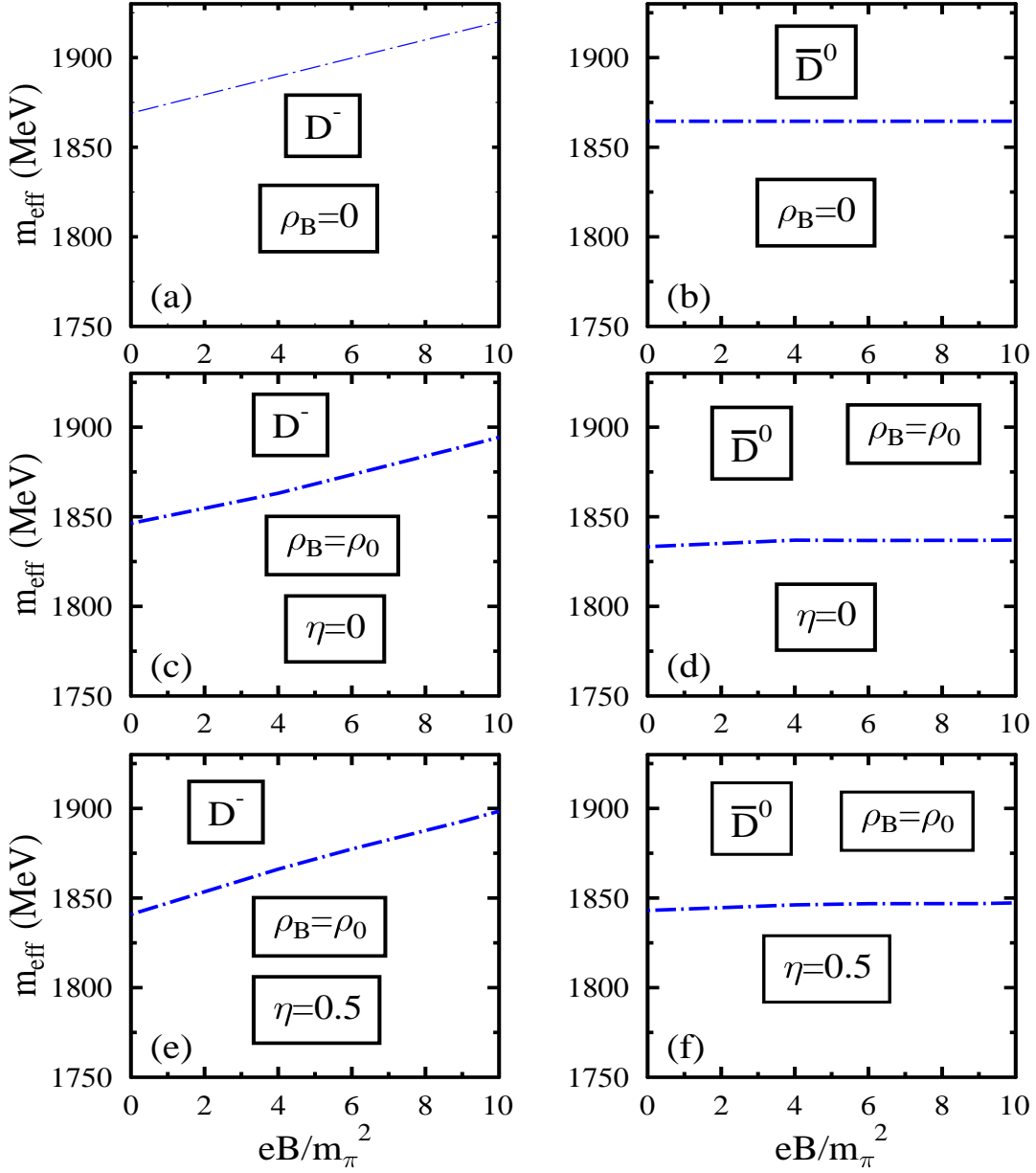}
\vskip -0.4in
\caption{(Color online)
Same as fig. \ref{dmeson_mag_rh0_f}, for $D^-$ and $\bar {D^0}$.
}
\label{dbar_mag_rh0_f}
\end{figure}

\begin{figure}
\hskip -0.4in
\vskip -0.4in
\includegraphics[width=18cm,height=18cm]{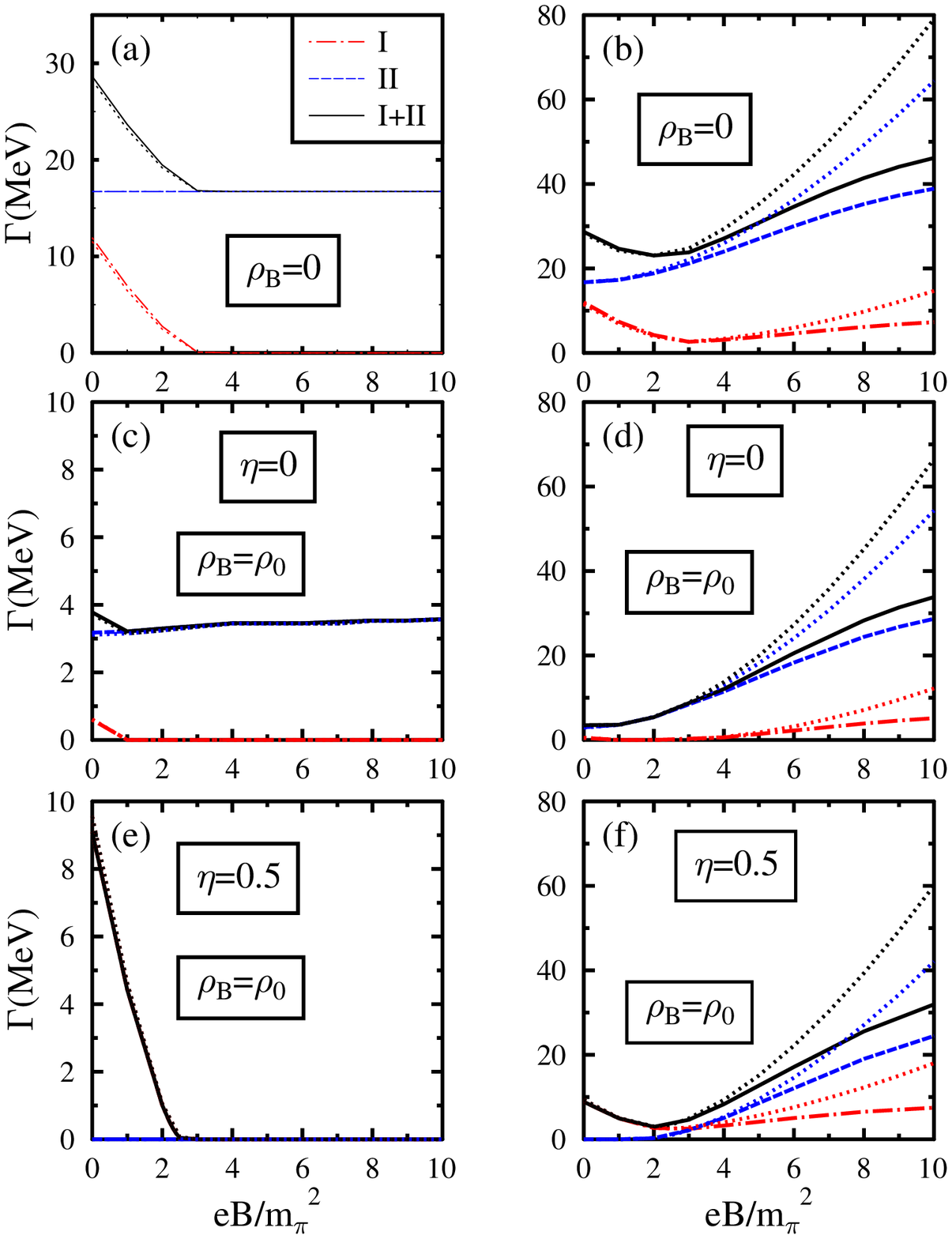}
\vskip -0.4in
\caption{(Color online)
Decay widths of $\psi(3770)$ to (I) $D^+D^-$,
(II) $D^0\bar {D^0}$, and the total of these two channels (I+II),
are plotted as functions of $eB/{m_\pi^2}$. These are shown 
for the case of $\rho_B=0$, as well as, for $\rho_B=\rho_0$,  
in symmetric ($\eta$=0) and asymmetric (with $\eta$=0.5) 
nuclear matter. The panels (a), (c) and (e) show the results
when the mixing with $\eta'_c$ is not taken into account,
and (b), (d) and (f) correspond to the results including the
mixing effect. These results are compared with the decay widths
obtained using a hadronic Lagrangian given by equation (\ref{eff_had}),
shown as dotted lines. 
}
\label{dwFT_3770_rh0_had}
\end{figure}

The in-medium masses of the vector charmonium states 
($J/\psi$, $\psi(2S)$, $\psi(1D)$) have been studied 
from the medium change of the gluon condensate,
using the leading order mass formula in Ref. \cite{leeko}
and using the chiral effective model in Refs. 
\cite{amarvdmesonTprc,amarvepja,charmonium_mag}
in (magnetized) hadronic matter. 
The mass modifications of the charmonium states
within the chiral effective model
arise from the modifications of a scalar 
dilaton field which simulates the gluon condensates
of QCD. These in-medium masses were studied 
assuming the harmonic oscillator wave functions
for the charm-anticharm bound states. The  values of the
strength parameter, $\beta \equiv 1/R$, of the charmonium 
wave functions
were fitted to the rms radii of $J/\psi$, $\psi(2S)$
and $\psi(1D)$ to be (0.47 fm)$^2$, (0.96 fm)$^2$ and 
1 fm$^2$ respectively \cite{leeko}, yielding their values 
as 513 MeV, 384 MeV and 368 MeV 
\cite{amarvdmesonTprc,amarvepja}. In the present work, 
the mass modifications for the pseudoscalar states,
$\eta_c\equiv \eta_c(1S)$ and $\eta_c' \equiv \eta_c (2S)$ 
in the nuclear matter are investigated using
the chiral effective model. 
The values of the harmonic oscillator
strengths for $\eta_c$ and $\eta_c'$ are obtained
as 535 MeV and 394.6 MeV respectively,
assuming these states to be lying in a straight line
with the states $J/\psi$ and $\psi(2S)$ in the mass 
versus $\beta$ graph.
In the absence of a magnetic feild, 
the masses for the pseudoscalar mesons
$\eta_c$ and $\eta_c'$ are modified to 2977.26 (2977.47)
and 3548.55 (3551.46) MeV at $\rho_B=\rho_0$
for symmetric (asymmetric with $\eta$=0.5) nuclear matter
from their vacuum masses of 2983.9 MeV and 3637.5 MeV 
respectively. The mass shift of $\eta_c$ 
in symmetric nuclear matter at the nuclear matter
saturation density of around 6.6 MeV  may be compared
to the values of 5 MeV \cite{klingl} and 
5.69 MeV \cite{amarvjpsi_qsr}
using a QCD sum rule approach, 
and of 3 MeV as calculated from $\eta_c$-nucleon scattering length
\cite{kaidalov_etac_mass}.
%%%%%%%%added below in response to comment 2 of Referee
The mass drop in $J/\psi$ of around 8.6 MeV at $\rho_B=\rho_0$
in symmetric nuclear matter obtained using the chiral effective model 
\cite{amarvdmesonTprc} 
may be compared with the mass drop of around 8 MeV calculated 
using second order Stark effect in QCD with the gluon condensate 
calculated in the linear density approximation \cite{leeko} 
as well as the mass shifts of 4 MeV and 7 MeV using QCD sum rule 
calculations with OPE upto dimension six and four
in Refs. \cite{kimlee} and \cite{klingl} respectively.
The mass shifts of the excited states $\psi(3686)$ and $\psi(3770)$ 
obtained in the chiral effective model 
\cite{amarvdmesonTprc,amarvepja} 
are observed to be similar to the values of
100 and 140 MeV using the QCD second order Stark effect  
\cite{leeko}, which are much larger
than the mass shift of $J/\psi$. 
%%%%%%added above

In figure \ref{mpsi_spinmix_eta0_f}, 
the effects of the magnetic field on the masses of the charmonium 
states are shown for zero baryon density as well as
for $\rho_B=\rho_0$ in symmetric ($\eta$=0) nuclear matter.
In panel (a), the masses of the pseudoscalar 
charmonium state, $\eta_c$ and the longitudinal component
of $J/\psi$ as modified due to the effect of mixing of 
these states, are shown as functions of $eB/m_\pi^2$. 
These masses are calculated by using the equation 
(\ref{mpv_long}). The parameter $g_{PV}
\equiv g_{ \eta_c J/\psi}$ is evaluated to be 2.094 from the observed 
radiative decay width $\Gamma (J/\psi \rightarrow \eta_c \gamma)$
in vacuum of 92.9 keV \cite{pdg_2019_update}, 
using equation (\ref{decay_VP}). 
The dotted lines correspond to the case
when these mixing effects are not taken into account.
For $\rho_B$=0, without the mixing effects,
the masses of the charmonium states 
are unaffected by the magnetic field, and remain at 
the vacuum values of the masses of $J/\psi$ and $\eta_c$  
of 3097 and 2983.9 MeV respectively.
%%%%%%%%%%%%%added below
The  mixing effect is observed to lead to appreciable
modifications to their masses at high magnetic fields, 
with an increase in the mass (in MeV) of $J/\psi^{||}$, 
the longitudinal component of $J/\psi$, by about
9 (31.5) and a drop in the mass of $\eta_c$ of around 
8.9 (30) at eB=5 (10) $m_\pi^2$ respectively.
The mass difference (in MeV) of $J/\psi$ and $\eta_c$ of 113 
in vacuum is thus modified to 131 (174.5) at value of 
the magnetic field $eB$ as 5 (10) $m_\pi^2$
for the $J/\psi^{||}$ and $\eta_c$ mesons.
Such a level repulsion of the $J/\psi^{||}$ and $\eta_c$
states has also been observed by incorporating the effects 
of a magnetic field consistently in the Operator product 
expansion (OPE) as well as in the phenomenological side 
within a QCD sum rule calculation
\cite{charmonium_mag_QSR,charmonium_mag_lee}.
An appropriate form of the spectral ansatz 
is used to describe the mixing 
of the pseudoscalar and longitudinal vector mesons,
retaining the effects of external magnetic field
upto second order in $eB$.
In Refs. \cite{charmonium_mag_QSR,charmonium_mag_lee},
the masses of the longitunal $J/\psi$ and $\eta_c$ 
in the presence a magnetic field calculated in the
QCD sum rule approach are compared with the masses 
from the effective hadronic interaction
\begin{equation}
{m_{P,V^{||}}^2}^{(PV)}_{\rm approx}
=m^2_{P,V}\mp \frac{{(g_{PV}eB)}^2}{m_V^2-m_P^2}, 
\label{m2pv_long_approx}
\end{equation}
obtained from the expressions given by (\ref{mpv_long})
by retaining terms upto the second order in $eB$ 
and the leading order in $\frac{(m_V-m_P)}{(m_V+m_P)}$ 
\cite{charmonium_mag_lee}. 
It is observed that the masses of the longitudinal $J/\psi$ 
and $\eta_c$ agree remarkably well with the second order
results of the masses obtained from the effective hadronic 
interaction upto $eB\sim 5 m_\pi^2
(\sim 0.1 GeV^2$) with slight deviation at higher values
of $eB$. Also, the second order results for the masses
of the $J/\psi^{||}$ and $\eta_c$ given by equation
(\ref{m2pv_long_approx}) are observed to 
agree extremely well with the general expressions
given by equation (\ref{mpv_long}) with marginal variations
at higher values of magnetic field \cite{charmonium_mag_lee}. 
In Refs. \cite{charmonium_mag_QSR,charmonium_mag_lee},
for comparing the results of the QCD sum rules with the
second order results of the effective hadronic model, 
the vacuum masses (in MeV) of $J/\psi$ and $\eta_c$ 
of 3092 and 3025 respectively in the QCD sum rule 
approach, are also taken as the vacuum masses 
for these mesons to compute the masses of $J/\psi^{||}$ and
$\eta_c$ due to their mixing.
One can see from the second order expression 
for the masses given by equation (\ref{m2pv_long_approx}) that
the mixing effect leads to an increase (drop)
in the mass of the longitudinal $J/\psi$ ($\eta_c$) meson.
The level repulsion of these states
was also studied using a potential model approach
by solving the Schrodinger equation in the presence
of an external magnetic field \cite{Alford_Strickland_2013}.
The masses of heavy quarkonium states have been studied
including the effects of mixing of the longitudinal
vector and pseudoscalar quarkonium states.
The Schrodinger equation for the heavy quark antiquark 
system is solved in the presence of a magnetic field,
assuming a Cornell potential along with spin-spin interaction.
In the presence of an external magnetic field,
the center of mass momentum of the quarkonium state is no longer
a conserved quantity. The masses of the lowest 
charmonium states $J/\psi$ and $\eta_c$ are obtained by solving
Schrodinger equation for given values of  $eB$
and the center of mass momentum, $\langle P_{kin}\rangle$
\cite{Alford_Strickland_2013}. For $\langle P_{kin}\rangle=0$,
there is observed to be an increase (drop) 
of $J/\psi^{||}$ ($\eta_c$) mass of the order of 80 (50) 
MeV for $eB=10 m_\pi^2 \sim 0.2 {\rm GeV}^2$ as is expected 
due to the spin-mixing \cite{Alford_Strickland_2013}. 
These values may be compared to the values of around 31.5 (30)
for the positive (negative) mass shifts of $J/\psi^{||}(\eta_c)$
in the present investigation. The mixing of the $\eta_c$
with the longitudinal $J/\psi$ can also have observational
consequences on the dilepton spectra. The probability of finding the
longitudinal $J/\psi$ in the state $\eta_c$ due to their mixing
has been calculated to be around 12 (26)\% for 
$eB\sim$ 5 (10) $m_\pi^2$ in Ref. \cite{Alford_Strickland_2013}.
This should show as a suppression of dileptons arising 
from the longitudinal $J/\psi$ (by about 12 (26)\%), 
which should instead show as decay from $\eta_c$ in the dilepton
spectra. The increase in the mass of $J/\psi^{||}$ in the presence
of strong magnetic fields should also show as a suppression 
in the $J/\psi$ production \cite{Alford_Strickland_2013}.

The masses of $J/\psi^{||}$ and $\eta_c$ are
plotted for $\rho_B=\rho_0$ in symmetric nuclear matter
in panel (b) in figure \ref{mpsi_spinmix_eta0_f}. 
In the absence of mixing effects, the masses 
are observed to vary marginally with magnetic field.
The mass difference (in MeV) of $J/\psi$ and $\eta_c$ of 
around 111 at $\rho_B=\rho_0$ without mixing effect,
is observed to be modified to 132 (175.7) for
the value of $eB$ as 5 (10) $m_\pi^2$ in the presence
of mixing for the longitudinal $J/\psi$ and $\eta_c$ mesons.
These values are observed to be similar to the values 
of the mass differences of these states of 
131 (174.5) MeV for $eB$=5 (10) $m_\pi^2$
for $\rho_B$=0.

The masses of $\psi ^{||}(2S)$ and $\eta'_c\equiv \eta_c (2S)$ 
in the presence of a magnetic field are shown 
for $\rho_B$=0 in panel (c) in figure \ref{mpsi_spinmix_eta0_f}. 
The value for the mixing coupling parameter of these states,
$g_{PV}\equiv g_{\eta'_c\psi(2S)}$ is obtained to be
3.184, fitted from the observed vacuum value of the decay width 
$\Gamma (\psi (2S) \rightarrow {\eta'_c} \gamma)$ of 0.2058 keV
 \cite{pdg_2019_update}. 
For $\rho_B=\rho_0$ as shown in panel (d), the
mixing effect is observed to be much more prominent
as compared to the zero density case. This is due to the
reason that the mass modification due to mixing is larger
for a smaller difference in the masses of the pseudoscalar 
and vector mesons. As can be seen from equation 
(\ref{m2pv_long_approx}),
the shifts in the masses of the pseudoscalar and vector
mesons are inversely proportional to $(m_V^2-m_P^2)$,
and hence inversely proportional to the mass difference
of these mesons. The mass splittings  (in MeV) 
of the $\psi^{||}(2S)$ and $\eta'_c$ are observed 
to be around 90.5 (174.9) for $eB$= 5 (10) $m_\pi^2$ 
at $\rho_B=\rho_0$ and 97 (174.88) for $\rho_B=0$ 
which are much larger as compared to the mass splitting 
of 48.5 in vacuum for zero magnetic field. As might be 
observed from the panels (c) and (d), the mass splitting 
for zero magnetic field is smaller for the $\rho_B=\rho_0$ 
case as compared to $\rho_B=0$.
However, the mass splitting is observed
to be similar for zero as well as nuclear matter saturation 
density for the considered magnetic fields, due to the larger
mass shifts for the longitudinal $\psi(2S)$ and $\eta'_c$
for $\rho_B=\rho_0$ compared to zero density.

The masses of $\eta'_c$ and the longitudinal component of
$\psi(1D)\equiv \psi(3770)$ in the presence of a magnetic field
are plotted for zero density and for $\rho_B=\rho_0$
in symmetric nuclear matter in panels (e) and (f) respectively 
in figure \ref{mpsi_spinmix_eta0_f}, including the
contributions from the mixing effects. The value for 
$g_{PV}\equiv g_{\eta'_c\psi(1D)}$ is obtained as
7.657 from the observed vacuum value of the decay width 
$\Gamma (\psi (2S) \rightarrow {\eta'_c} \gamma)$
of 24.48 keV \cite{pdg_2019_update}. 
Due to the smaller difference
in the masses of the $\eta'_c$ and $\psi(3770)$
in the magnetized nuclear matter, the mixing effects 
are observed to lead to larger shifts in the masses 
of these mesons at $\rho_B=\rho_0$ as compared to zero density.
This is similar to the larger mass shifts of $\eta_c'$ 
and $\psi(3686)$ due to the mixing effects,
at $\rho_B=\rho_0$ as compared to zero density. 
The mass splittings of the $\psi^{||}(1D)$
and $\eta'_c$ are observed to be around  188.2 (332.7)
for $eB$= 5(10) $m_\pi^2$ at $\rho_B=\rho_0$ and 
213.7 (346.2) for $\rho_B=0$, which is much larger 
than the mass difference of 135.5 in vacuum for zero magnetic
field. As can be observed from the figure, 
the mass of the longitudinal $\psi(3770)$ 
has dominant positive contributions at high magnetic fields 
due to the mixing effects.
This leads to significant modification of the partial decay width
of $\psi(3770)$ to $D\bar D$ at high magnetic fields, 
as we shall see later.

The mixing of the longitudinal component of the vector charmonium state
with the pseudoscalar mesons can lead to 
`anomalous' $\eta_c$ and $\eta'_c$ peaks,
in addition to the $J/\psi$, $\psi(3686)$ and $\psi(3770)$-like peaks 
in the dilepton spectra, and can probe the existence 
of early magnetic field. 
A study of the formation times of the charmonium states
due to the mixing effects of $J/\psi-\eta_c$ as well as 
$\psi(3686)-\eta'_c$ is observed to lead to a faster (slower)
formation of the pseudoscalar (vector) charmonium state.
The early formation of the $\eta_c$ and $\eta'_c$ can
probe the magnetic field at the early stage when 
the magnetic field can still be large.

The charmonium masses are plotted for asymmetric nuclear
matter with $\eta$=0.5 in figure \ref{mpsi_spinmix_eta5_f}.
The effects of the isospin asymmetry on the
charmonium masses are observed to be
small at the nuclear matter saturation density, 
as calculated within the chiral effective
model. This, in turn, leads to small modifications
to the charmonium masses in the isospin asymmetric 
matter (with $\eta$=0.5) as compared to the symmetric 
nuclear matter, in the presence of mixing effects. 
For example, the masses (in MeV) of $\psi(3686)$ 
and $\psi(3770)$ at the nuclear matter saturation density, 
and for $eB=10 m_\pi^2$, the highest magnetic field
considered in the present work,
are observed to be modified from 3650.57 and 3808.45 for symmetric 
nuclear matter to the values 3662.2 and 3821.4 for asymmetric 
nuclear matter with $\eta$=0.5. 
It might be noted here that the contributions 
to the masses of the excited charmonium states 
$\psi(3686)$ and $\psi(3770)$, at zero density
as well as at $\rho_B=\rho_0$ in nuclear matter, 
due to the mixing with $\eta_c$ are negligible 
for the magnetic fields considered in the present work. 
This is expected from the larger mass differences
of these vector mesons from the mass of $\eta_c$,
which give negligible shifts to these charmonium
masses due to mixing with $\eta_c$, as is evident
from equation (\ref{m2pv_long_approx}). 
The excited states $\psi(3686)$ and $\psi(3770)$
thus have dominant contributions to their masses
arising from the mixing with $\eta'_c$ in the
presence of high magnetic fields, as has been
discussed above.

The mass splittings of the longitudinal component
of the vector (V$^{||}$) and the pseudoscalar (P) charmonium states
are observed to much larger for high magnetic fields
than the values in vacuum at zero magnetif field, due to the
effects of mixing. For $J/\psi^{||}-\eta_c$, $\psi^{||}(3686)-\eta'_c$
and $\psi^{||}(3770)-\eta'_c$ mixings, the values of mass
difference in these states (in MeV) of 113, 48.5 and 135.5
for vacuum at zero magnetic field are modified to
around 130 (175), 90 (175) and 188 (333) for $eB=5 (10)m_\pi^2$.
These large mass splittings should show as peaks corresponding
to these states in the dilepton spectra in ultrarelativistic
heavy ion collision experiments and can probe
the existence of magnetic field in the early stage. 

In figures \ref{dmeson_mag_rh0_f} and  \ref{dbar_mag_rh0_f},
the masses of the $D(D^+,D^0)$ mesons and $\bar D(D^-,\bar {D^0})$ 
mesons, are plotted as functions of $eB/m_\pi^2$,
for $\rho_B=0$ as well as for $\rho_B=\rho_0$ with $\eta$=0
and $\eta=0.5$. In the nuclear matter in the presence
of a magnetic field, these masses are calculated within the chiral 
effective model, from their interactions with the nucleons
and scalar mesons. The charged open charm mesons ($D^+$ and $D^-$) 
have additional positive shifts in their masses due to contributions 
from the Landau levels in the presence of a magnetic field. 
On the other hand, the neutral $D^0$ and $\bar {D^0}$ mesons 
are observed to have very small changes in their masses due 
to the magnetic field. The $D$ and $\bar D$ masses in nuclear matter
have been calculated in presence of magnetic fields 
using the chiral effective model in Ref. \cite{dmeson_mag},
without accounting for the medium modification of the
scalar dilaton field, which simulates the gluon condensates 
of QCD. Using the chiral effective model, the mass modifications
of the charmonium states arise due to the medium changes
of the dilaton field. The in-medium decay widths of the charmonium
states to $D\bar D$ have been studied using the mass modifications
of the charmonium states \cite{charmonium_mag} 
as well as the open charm mesons in the magnetized nuclear matter
\cite{charm_decay_mag_3p0}. In the present work, the
vector charmonium masses have been studied including the
mixing of the longitudinal component of these states
with the pseudoscalar charmonium states.
The mass of the longitudinal component of
the vector charmonium state has an appreciable positive shift 
due to the pseudoscalar--vector meson mixing in the
presence of the magnetic field.
The lowest charmonium state decaying to $D\bar D$
is $\psi(3770)$ at $\rho_B=\rho_0$ as well as at
zero baryon density.
As has already been mentioned, the charmonium decay width, 
$\Gamma (\psi(3770) \rightarrow D\bar D)$
is calculated using the light quark
pair creation term of the free Dirac Hamiltonian 
expressed in terms of the constituent quark
operators, using explicit constructions
for the charmonium and the  open charm 
($D$ and $\bar D$) mesons. 
The matrix element for the calculation
of the decay width is multiplied with a factor
$\gamma_\psi$, which gives the strength of the
light quark pair creation leading to the decay
of the charmonium state to $D\bar D$ in the 
magnetized hadronic medium \cite{amspmwg}. 
The value of $\gamma_\psi$
to be 1.35, is chosen so as to reproduce the decay widths
of $\psi(3770) \rightarrow D^+ D^-$ and   
$\psi(3770) \rightarrow D^0 \bar{D^0}$ in vacuum,
to be around 12 MeV and 16 MeV respectively 
\cite{amspmwg}. The constituent quark masses
for the light quarks ($u$ and $d$) are taken to 
be 330 MeV and for the charm quark, the value
is taken to be $M_c=1600$ MeV \cite{amspmwg}.

The partial decay widths of $\psi(3770)$ to 
$D\bar D(D^+D^-,D^0 \bar {D^0})$,
are calculated using a field theoretic model for composite
hadrons as described in section \ref{dwFT_comp_had}.
These are plotted in figure \ref{dwFT_3770_rh0_had}
as functions of $eB/m_\pi^2$ for $\rho_B=\rho_0$ 
as well as at zero baryon density.
%%%%%%added below
To examine the effects of the compositeness
of the hadrons on the charmonium decay widths in the 
presence of strong magnetic fields,
the decay widths of $\psi(3770)\rightarrow D\bar D$ as
calculated using a field theoretic model of hadrons with
quark/antiquark constituents are compared with the results
obtained using the effective hadronic Lagrangian \cite{HGA} 
given by equation (\ref{eff_had}) (shown as the dotted lines).
The value of the coupling constant $g_{\psi D\bar D}$
in the interaction hadronic Lagrangian
is taken as 14.5, which gives the vacuum values
of the decay widths of $\psi (3770)\rightarrow D \bar D$
to be around 28 MeV, with the values of 12 and 16 MeV in the decay 
channels to $D^+D^-$ and $D^0 {\bar {D^0}}$ respectively.
%%%%%%added above
In panel (a) in figure \ref{dwFT_3770_rh0_had},
the partial decay widths of $\psi(3770)$ to (I) $D^+D^-$,
(II) $D^0\bar {D^0}$, and the total of these two channels (I+II),
are plotted for $\rho_B=0$, without taking into account 
the pseudoscalar--vector meson ($\eta'_c-\psi(3770)$) mixing.
The decay width for $\psi(3770)$ decaying to the charged $D\bar D$ 
is observed to drop with increase in the magnetic field upto 
$eB$ equal to $3m_\pi^2$, when it becomes zero and remains zero
for larger values of the magnetic field. This is due to
the reason that the charged $D$ and $\bar D$ mesons
have  higher values for their masses  
in the presence of a magnetic field, due to 
positive Landau level contributions to their masses.
On the other hand, the decay wdith to the neutral $D\bar D$ pair 
is observed to be unaffected by the magnetic field
in the absence of the mixing effects as seen in panel (a).
Panel (b) shows the partial decay widths when the
mass modification of the longitudinal component of
$\psi(3770)$ is taken into account due to mixing with
$\eta'_c$ in the presence of a magnetic field.
This leads to an appreciable increase in the decay to neutral
$D^0 \bar {D^0}$ with rise in the magnetic field,
whereas the decay to the $D^+D^-$ is observed to have
an initial drop upto around $eB\sim 3 m_\pi^2$ 
beyond which it is observed to be almost constant 
(with a value of around 7 MeV) when calculated using the field 
theoretic model of composite hadrons.
There are observed to be appreciable deviations 
in the decay widths computed using the  model
for composite hadrons from the calculations within the effective 
hadronic model at high values of the magnetic fields
in the presence of mixing effects.
As can be seen from the
equations (\ref{gammapsiddbar}) 
and (\ref{gammapsippddbar_had}), 
the dependence of the charmonium decay widths to $D\bar D$
on the magnitude of the momentum of the outgoing 
are given as a polynomial multiplied by an exponential
function in the model of  composite hadrons, whereas,
there is a $|{\bf p}|^3$ dependence
in the effective hadronic model. 
These are modified to the expressions given 
by equations (\ref{gammapsiddbar_mix})
and (\ref{gammapsiddbar_mix_had}) in the presence of magnetic 
fields, as the longitudinal components of the vector charmonium states
are modified due to mixing with the pseudoscalar charmonium states.
The in-medium decay widths of charmonium states to $D\bar D$
have been studied due to mass modifications of the
open charm mesons in hadronic matter for zero magnetic field,
using the compositeness of the hadrons using the $^3P_0$ model
\cite{friman,amarvepja} as well as using the field theoretic
model for composite hadrons \cite{amspmwg} as considered 
in the present work, are observed to vanish at certain
densities (so called nodes). This is due to the dependence 
of these decay widths on the magnitude of the momentum 
of the outgoing $D(\bar D)$ meson, $|{\bf }|$, as a polynomial
multiplied by an exponential function, which vanishes
for the values of the $|{\bf p}|$ corresponding to
these densities. On the other hand,
the effective hadronic model as given by (\ref{eff_had}),
due to the $|{\bf p}|^3$ as given by equation 
(\ref{gammapsippddbar_had})
is observed to lead to a monotonic increase with increase
in $|{\bf p}|$ (corresponding to smaller values of the 
$D(\bar D)$ mass in the medium). The results
for the charmonium decay widths obtained
from the effective hadronic model are significantly 
different from the behaviour as obtained in the $^3P_0$ model
\cite{friman}. In the present work, the increase in the
value of $|{\bf p}|$ in the presence of magnetic fields
is due to the appreciable increase in the mass
of the longitudinal component of the charmonium state, 
$\psi(3770)$ due to mixing with the pseudoscalar meson, 
$\eta'_c$, which is observed to lead to much larger increase
in the decay width of $\psi(3770)\rightarrow D^0\bar {D^0}$
obtained from the effective hadronic model,
as compared to the values calculated in the field theoretic 
model. The presence of a magnetic field leads to increase
in the masses of the $D^\pm$ mesons due to the Landau level
contributions, which is observed as a drop in the decay width
upto the value of $eB\sim 3 m_\pi^2$, beyond which it is
observed to remain almost constant (increases) with further
increase in the magnetic field, in the model for composite
hadrons (effective hadronic model) as can be seen in panel (b)
for zero baryon density and including the  effect of mixing
of the longitudinal $\psi(3770)$ with $\eta'_c$.

The decay widths of $\psi(3770)\rightarrow D\bar D$
for baryon density $\rho_B=\rho_0$
in magnetized nuclear matter are shown  
without and with the mixing effects, 
in panels (c) and (d) for isospin asymmetry
parameter, $\eta$=0 and in panels (e) and (f) 
for $\eta$=0.5 in figure \ref{dwFT_3770_rh0_had}. 
In the absence of mixing effects, these decay widths 
are obtained from the modifications of the masses of 
the charmonium state $\psi(3770)$ as well as 
the $D$ and $\bar D$ mesons in magnetized nuclear
matter using the chiral effective model. 
In the absence of mixing effects, the decay width
to charged $D^+D^-$ is observed to be around
0.5 MeV at zero magnetic field and vanishes
for $eB$ higher than $m_\pi^2$, when the drop
in the charmonium mass in the medium as well as
the increase in the masses of the charged open charm mesons
due to the Landau contributions in the presence of the magnetic 
field kinematically forbids the decay of $\psi(3770)$ to $D^+D^-$. 
The decay to the neutral $D\bar D$ is observed to
increase with magnetic field and the rise is much higher 
in the effective hadronic model due to the $|{\bf p}|^3$
dependence of the decay width. 
The panels (e) and (f) show the results
for the decay widths for asymmetric nuclear matter
with $\eta$=0.5 for $\rho_B=\rho_0$.
In the absence of mixing effects, there is observed to be
contribution to the decay width of $\psi(3770)$
only from the channel of decay to the charged $D\bar D$ pair.
The increase in the masses of the final state
charged open charm mesons with magnetic field
due to the Landau contributions is observed as
a drop and vanishing of the decay wdith 
of $\psi(3770)\rightarrow D^+D^-$
for $eB$ higher than around 2.3 $m_\pi^2$. 
In the presence of the mixing effects, due to increase
in the mass of the longitudinal $\psi(3770)$,
there is observed to be contribution 
from the partial decay width to the
neutral open charm meson pair for $eB$
higher than 2 $m_\pi^2$.
For larger values of the magnetic field,
the decay width is observed to be dominated by
the contribution from the neutral $D\bar D$ channel.
However, the decay widths obtained within the
effective hadronic model, shown as dotted lines,
are observed to have a much sharper rise as compared
to the results obtained using the  model of composite
hadrons. 

The decay of the charmonium state $\psi(3770)\rightarrow D\bar D$ 
is observed to be dominated by the decay to the neutral
$D\bar D$ pair as compared to $D^+D^-$, for large 
values of the magnetic fields, both for the zero baryon density,
as well as for $\rho_B=\rho_0$ in (asymmetric) nuclear matter.
This should lead to the neutral open charm mesons ($D^0$
and $\bar {D^0}$) to be more abundant as compared to the charged 
$D^\pm$ mesons. The decay width is observed to increase
with the magnetic field. This should lead to suppression
in the yield of $\psi(3770)$ at large values of the magnetic 
fields, hence of the $J/\psi$, as the excited charmonium 
states are a major source of $J/\psi$.

\section{Summary}
In the present work, the charmonium masses as well as the
decay widths of the vector charmonium states to $D\bar D$
are investigated in the presence of strong magnetic fields.
The study can be of relevance to the observables
in ultra-relativistic heavy ion collision experiments.
The estimated magnetic fields created in the peripheral 
heavy ion collisions, e.g., at RHIC and at LHC,
are huge, where the formed matter is (extremely) low in density.
In the present work, we study the effects of the
magnetic fields on the masses and decay widths
of the charmonium states at zero as well as 
at nuclear matter saturation density.

The mixing of the pseudoscalar mesons and the vector mesons 
in the presence of strong magnetic fields are studied using an
effective Lagrangian interaction. The effects from the
mixing are observed to have dominant 
contributions to the masses of the charmonium states,
with an increase (drop) for the longitudinal component
of the vector (pseudoscalar) charmonium state
in vacuum/nuclear matter.
In the nuclear medium, the masses of the vector charmonium states
($J/\psi$, $\psi(3686)$, $\psi(3770)$) and 
the pseudoscalar states $\eta_c$ and $\eta'_c$ are
studied as arising from the medium changes of the 
scalar dilaton field (which simulates the gluon
condensates of QCD) using a chiral effective model. 
These masses charmonium states are additionally modified 
in the nuclear medium due to the mixing effects.
The mixing of the pseudoscalar mesons, $\eta_c$ and
$\eta'_c$ with the longitudinal vector (V) charmonium states,
lead to large positive (negative) shifts
in the masses of $V^{||}$ (pseudoscalar) mesons at high 
magnetic fields. The finite probability of the longitudinal
component of the vector meson to be in the pseudoscalar 
state should lead to suppression of the dileptons from the
decay of $V^{||}$, which should instead arise from the
decay of the pseudoscalar meson. The mixing effect can thus
show as peaks in the dilepton spectra
due to `anomalous' decay modes, e.g.,
$\eta_c,\eta_c'\rightarrow l^+l^-$ in addition
to the $V\rightarrow l^+l^-$ and can probe the
magnetic field at the early stage. 

The in-medium decay widths of charmonium state to $D\bar D$
in the presence of strong magnetic fields
are computed using a field theoretical model for composite hadrons
with quark (antiquark) constituents. 
The matrix element for computation of the charmonium 
decay width is calculated from the free Dirac
Hamiltonian of the constituent quarks, using
the explicit constructions of the charmonium 
state, the $D$ and the ${\bar D}$ mesons.
The decay widths are computed from the modifications
of the masses of the charmonium state and the open charm
mesons in vacuum/nuclear medium in the presence
of strong magnetic fields. In the nuclear medium, 
the charmonium masses as well as the $D$ and $\bar D$ masses
are calculated within a chiral effective model.
As has already been mentioned, the charmonium masses
are calculated from the medium modification
of a scalar dilaton field, and, the masses of the $D$ and $\bar D$ mesons
are calculated within the model due to their interactions
with the nucleons and the scalar mesons in the magnetized
nuclear matter, with additonal Landau contributions to the masses
of the charged $D^\pm$ mesons in the presence of the magnetic
field. At $\rho_B$=0  as well as at nuclear matter saturation density,
the lowest charmonium state which can decay to $D\bar D$ 
is $\psi(3770)$. 
The mass of longitudinal $\psi(3770)$ is observed to have
appreciable positive contribution due to mixing with $\eta_c'$ 
in the presence of strong magnetic fields, 
which is seen to have significant
modification to the partial decay width, 
$\Gamma (\psi (3770)\rightarrow D\bar D)$.
The decay of $\psi(3770)$ to $D^+D^-$ are observed to be
quite suppressed as compared to decay to $D^0 \bar {D^0}$,
in the presence of strong magnetic fields. This is  
due to the higher masses of the charged mesons ($D^+$,$D^-$)
due to contributions from the Landau levels. This should lead to 
the production of the neutral $D^0$ and $\bar {D^0}$ 
mesons to be more abundant as compared to the 
$D^\pm$ mesons from $\psi(3770)$. The increase of
the decay width of $\psi(3770)$  
at high magnetic fields should show as a suppression 
in the yield of $\psi(3770)$ and hence of $J/\psi$,
as $\psi(3770)$ is a major source for production of $J/\psi$.

To summarize, the present study of the masses of the charmonium states in 
strong magnetic fields, which has dominant contribtuions
from the mixing of the longitudinal vector meson -- pseudoscalar mesons
can be a probe for the early magnetic field as the mixing can
show as pseudoscalar like peaks in addition to the vector meson peaks
in the dilepton spectra, and the study of charmonium 
decay width to $D\bar D$, which is observed to be enhanced 
in strong magnetic fields, predominantly in the decay mode 
to neutral $D\bar D$, should have observable consequences 
on the suppression in the yield of $\psi(3770)$ (and hence of $J/\psi$)
and higher yield for the neutral open charm ($D^0$, $\bar {D^0}$) 
as compared to the charged $D^\pm$ mesons due to the presence
of strong magnetic fields created in non-central ultra-relativistic 
heavy ion collision experiments.

One of the authors (AM) is grateful to ITP, University of Frankfurt,
for warm hospitality and 
acknowledges financial support from Alexander von Humboldt Stiftung 
when this work was initiated. 

%\begin{references}

%\end{references}
\end{document}